\documentclass{iopart}
\usepackage{iopams}
\begin{document}

\title[A Perturbative Gravitational Wave in Quadratic Gravity]
{A Perturbative Solution for Gravitational Waves in Quadratic
Gravity}

\author[de Rey Neto]{Edgard C. de Rey Neto, Odylio D. Aguiar, Jos\'e C. N. de
Araujo}

\address{Instituto Nacional de Pesquisas Espaciais - Divis\~ao de
Astrof\'\i sica \\ Av. dos Astronautas 1758, S\~ao Jos\'e dos
Campos, 12227-010 SP, Brazil}

\ead{\mailto{edgard@das.inpe.br},
     \mailto{odylio@das.inpe.br}, and
     \mailto{jcarlos@das.inpe.br}}

\begin{abstract}
We find a gravitational wave solution to the linearized version of
quadratic gravity by adding successive perturbations to the
Einstein's linearized field equations. We show that only the Ricci
squared quadratic invariant contributes to give a different
solution of those found in Einstein's general relativity. The
perturbative solution is written as a power series in the $\beta$
parameter, the coefficient of the Ricci squared term in the
quadratic gravitational action. We also show that, for
monochromatic waves of a given angular frequency $\omega$, the
perturbative solution can be summed out to give an exact solution
to linearized version of quadratic gravity, for
$0<\omega<c/\!\mid\!\!\beta\!\!\mid^{1/2}$.
This result may lead to implications to the predictions for gravitational wave backgrounds of cosmological origin.
\end{abstract}

\submitto{\CQG}


\maketitle

\newpage

\section{Introduction}

Gravitational Lagrangians which include quadratic curvature terms,
have been proposed at classical level as an extension of
Einstein's theory~\cite{weyl}. More recently, quadratic
Lagrangians have been used to yield renormalizable theories of
gravity coupled to matter~\cite{stelle1}. Higher derivative
theories also arise as the low energy limit of string
theories~\cite{zwiebach,tseytlin}.

For the sake of definiteness we will deal with the quadratic (fourth-order)
gravity generated by the action

\begin{eqnarray}
\label{ac1}
S=\frac{1}{16\pi G}{\int d^4x\sqrt{-g}\left  \{R+\alpha  R^2+\beta
R_{\mu\nu}R^{\mu\nu}+16\pi G{\cal L}_m\right\}}.
\end{eqnarray}
Here, $R$ is the Ricci scalar, $R_{\mu\nu}$ is the Ricci
tensor, $G$ is the Newton's gravitational constant, and $\alpha $,
$\beta$ the coupling parameters of the quadratic interactions. We
use throughout this work units in which $\hbar=c=1$, metric
signature $(+---)$, Riemann tensor ${R^\alpha }_{\beta
{\mu\nu}}=-\Gamma ^\alpha _{\beta \mu;\nu}+\cdots$ and Ricci
tensor $R_{\mu\nu}= g^{\alpha \beta }R_{\alpha {\mu\nu}\beta }$.

Let us consider the action $S$. For $\alpha =\beta =0$, $S$ is
exactly reduced to the Einstein-Hilbert action. Then, we can
expect that for $\alpha $ and $\beta $ sufficiently small the
quadratic gravity can be approximately described by Einstein's
General Relativity second-order equations with small corrections
due to quadratic curvature couplings. In this case approximate
solutions to quadratic gravity can be constructed by adding
successive perturbations to a solution of the Einstein's gravity.
This perturbative approach is develop from the concept of
regular reduction of a system of partial differential
equations~\cite{belzia} and is used by Lousto {\it et
al.}~\cite{pmt1,pmtbh,comen} to obtain perturbative solutions to
quadratic gravity for some non-radiative astrophysical scenarios.
In the present work instead, we use a perturbative approach to
build radiative solutions to quadratic gravity.

In section~\ref{sec2}, we give a description of the scheme on how
the perturbative field equations are obtained. In
section~\ref{sec3}, we use the perturbative method to build a
solution to the linearized quadratic gravity.
In section~\ref{sec4} we discuss the implications of the solution obtained to astrophysics and add some considerations concerning the detectability of the computed effect.

\section{The general perturbative scheme}
\label{sec2}

We give in this section, a description of the general scheme on
how the perturbative equations are obtained. We then show how a
solution to quadratic gravity, as a power series in $\alpha $ and
$\beta $ parameters, can be built.

Let us now write the general field equations to the fourth-order
gravity. Variations of the action $S$, with respect to the metric
$g_{\mu\nu}$ leads to the following field equations:

\begin{eqnarray}
\label{eulageq1}
G_{\mu\nu}+\alpha  H_{\mu\nu}+\beta  I_{\mu\nu}=-8\pi GT_{\mu\nu},
\end{eqnarray}

\noindent where $G_{\mu\nu}$ is the Einstein's tensor given by

\begin{eqnarray}
\label{Gdef}
G_{\mu\nu}=R_{\mu\nu}-\frac{1}{2} g_{\mu\nu} R.
\end{eqnarray}

The terms in the fourth-order derivatives of the space-time metric
are given by

\begin{eqnarray}
\label{Hdef}
H_{\mu\nu}=2R_{,{\mu\nu}}-2g_{\mu\nu}\opensquare R+\frac{1}{2} g_{\mu\nu}
R^2-2RR_{\mu\nu}
\end{eqnarray}

\noindent and

\begin{eqnarray}
\label{Idef}
I_{\mu\nu}=R_{,{\mu\nu}}-\frac{1}{2} g_{\mu\nu}\opensquare R-\opensquare
R_{\mu\nu}-2{R_\mu}^{\alpha }R_{\alpha \nu}+\frac{1}{2} g_{\mu\nu}
R_{\alpha \beta }R^{\alpha \beta }.
\end{eqnarray}

Let us now write the perturbative equations for the
system~(\ref{eulageq1}), that is, the equations which give as
solutions the contribution to the space-time metric of a given
order in $\alpha $ and $\beta $ parameters.

In the {\it zeroth-order} ($\alpha =\beta =0$), the perturbative
equation is by construction the ordinary Einstein's gravity
equation, which is written as

\begin{eqnarray}
\label{eqo0}
G^{(0)}_{\mu\nu}=-8\pi GT^{(0)}_{\mu\nu},
\end{eqnarray}

\noindent where

\begin{eqnarray}
G^{(0)}_{\mu\nu}=R^{(0)}_{\mu\nu}-\frac{1}{2} g_{\mu\nu} R^{(0)}.
\end{eqnarray}

The solution of these equations is denoted by $g^{(0)}_{\mu\nu}$
and is independent of $\alpha $ and $\beta $ parameters. From the
above relations we obtain

\begin{eqnarray}
\label{r0}
R^{(0)}=8\pi GT^{(0)}
\end{eqnarray}

\noindent and

\label{rmn0}
\begin{eqnarray}
R^{(0)}_{\mu\nu}=-8\pi G\left  (T^{(0)}_{\mu\nu}-\frac{1}{2}\eta_{\mu\nu}
T^{(0)}\right) .
\end{eqnarray}

At the {\it 1th-order} we have the equations

\begin{eqnarray}
\label{eqo1}
G^{(1)}_{\mu\nu}=-8\pi GT^{(1)}_{\mu\nu}-\alpha  H^{(0)}_{\mu\nu}-\beta
I^{(0)}_{\mu\nu},
\end{eqnarray}

\noindent where $H^{(0)}_{\mu\nu}$ and $I^{(0)}_{\mu\nu}$ can be
written in terms of the energy-momentum tensor from~(\ref{r0})
and~(\ref{rmn0}). The solution of these equations at first order
in the coupling parameters $\alpha $ and $\beta$ is denoted by
$g^{(1)}_{\mu\nu}$.

By repeating this procedure to higher order we obtain at {\it
nth-order} the following of perturbative equations:

\begin{eqnarray}
\label{eqon}
G^{(n)}_{\mu\nu}=-8\pi G T^{(n)}_{\mu\nu}-\alpha  H^{(n-1)}_{\mu\nu}-\beta
I^{(n-1)}_{\mu\nu}.
\end{eqnarray}

These equations show that at each order of perturbation scheme we
have the Einstein's General Relativity equations with a new
effective energy-momentum source. Thus, in this perturbative
scheme, the quadratic curvature couplings are left out of the
dynamics, which is given by Einstein's gravity equation. The
higher order terms enter only as components of the source in the
perturbative field equations.

The approximate solution to the quadratic gravity can be
constructed by adding the contributions to the space-time metric
order by order, that is,

\begin{eqnarray}
\label{gserab}
g_{\mu\nu}=g^{(0)}_{\mu\nu}+g^{(1)}_{\mu\nu}+g^{(2)}_{\mu\nu}+\cdots.
\end{eqnarray}

\section{The perturbative scheme in linear approximation}
\label{sec3}

The ordinary theory of gravitational emission and propagation is
developed in the radiation zone, where the variations of the
gravitational field can be treated linearly. Then, we apply the
scheme delineated above to the linearized quadratic gravity. To
deal with the linearized theory we expand the metric $g_{\mu\nu}$
as follows

\begin{eqnarray}
\label{linfielddef}
g_{\mu\nu}=\eta_{\mu\nu}+\kappa h_{\mu\nu},
\end{eqnarray}

\noindent where $\kappa\ll1$, and $\eta_{\mu\nu}$ is the Minkowski
space-time metric. Within this approximation, we can neglect the
nonlinear curvature terms present in~(\ref{Hdef})
and~(\ref{Idef}). Thus, the linearized version of the
eq.~(\ref{eulageq1}) can be written as

\begin{eqnarray}
\label{linfulleq}
G^{(L)}_{\mu\nu}=-\kappa T_{\mu\nu}-2\frac{\alpha
}{\kappa}H^{(L)}_{\mu\nu}-2\frac{\beta}{\kappa}I^{(L)}_{\mu\nu},
\end{eqnarray}

\noindent where $\kappa=\sqrt{16\pi G}$, $G^{(L)}_{\mu\nu}$ is the
linearized Einstein tensor defined by

\begin{eqnarray}
G^{(L)}_{\mu\nu}\equiv\opensquare
h_{{\mu\nu}}-h_{(\mu\alpha \;\;\;\;\nu)}^{\;\;\;\;\;\;,\alpha
}+h_{,{\mu\nu}}-\eta_{\mu\nu}(\opensquare
h-{h^{\alpha \beta }}_{,\alpha \beta  })
\end{eqnarray}

\noindent and the terms which involve fourth-order derivatives of
the metric reduce after linearization to

\begin{eqnarray}
\frac{H^{(L)}_{\mu\nu}}{\kappa}=2\opensquare h_{,{\mu\nu}}-2{h^{\alpha \beta
}}_{,\alpha \beta {\mu\nu}}-2\eta_{\mu\nu}
(\opensquare^2h-\opensquare {h^{\alpha \beta }}_{,\alpha \beta })
\end{eqnarray}

\noindent and

\begin{eqnarray}
\eqalign{
\frac{I^{(L)}_{\mu\nu}}{\kappa}&=\opensquare h_{,{\mu\nu}}-{h^{\alpha \beta
}}_{,\alpha \beta {\mu\nu}}-\frac{1}{2}\eta_{\mu\nu}\opensquare
(\opensquare h-{h^{\alpha \beta }}_{,\alpha \beta })  \\
&-\frac{1}{2}\opensquare\left  (\opensquare h_{\mu\nu}-
h_{(\mu\alpha \;\;\;\;\nu)}^{\;\;\;\;\;\;,\alpha }+h_{,{\mu\nu}}\right) .}
\end{eqnarray}

Let us adopt the Hilbert gauge, namely

\begin{eqnarray}
\label{higauge}
h_{\mu\nu}^{\;\;\;\;,\nu}=\frac{1}{2} h_{,\mu}
\end{eqnarray}

\noindent and define the trace reverse of $h_{\mu\nu}$ by

\begin{eqnarray}
\label{tracerev}
\varphi_{\mu\nu}\equiv h_{\mu\nu}-\frac{1}{2}\eta_{\mu\nu} h.
\end{eqnarray}

In our approximation we deal with metric independent sources,
therefore the energy-momentum tensor is independent of $\alpha $
and $\beta $ parameters. Thus,

\begin{equation}
\label{tmno}
\eqalign{
T^{(0)}_{\mu\nu}=T_{\mu\nu}\qquad{\rm and}\qquad\\
T^{(n)}_{\mu\nu}=0\qquad{\rm for}\qquad n\geq1.}
\end{equation}

Now, we write the perturbative equations expanding the field
variable $\varphi_{\mu\nu}$, following the scheme delineated in
the previous section.

The {\it zeroth-order} perturbative equation is the linearized Einstein's
gravity equation

\begin{eqnarray}
\label{eqlin0}
\opensquare \varphi^{(0)}_{\mu\nu}=-\kappa T_{\mu\nu}^{(0)},
\end{eqnarray}

\noindent whose solutions in the radiation zone are given by the
usual retarded gravitational potential

\begin{eqnarray}
\varphi^{(0)}_{\mu\nu}=-\frac{\kappa}{4\pi r}{\int d^3{\bf x}^\prime
T_{\mu\nu}(t-r,{\bf x}^\prime)},
\end{eqnarray}

\noindent where $r$ denotes the distance from the source to the
observer which is very large as compared to the dimensions of the
source.

These are just the ordinary Einstein's General Relativity solutions from which
we know that they can be made transverse, traceless and have only spatial
components.

By virtue of~(\ref{tmno}), and the traceless nature of
$\varphi^{(0)}_{kl}$ the {\it 1-th order} perturbative equations
are independent of $\alpha $ and reduce to

\begin{eqnarray}
\label{eqlin1}
\opensquare\varphi^{(1)}_{\mu\nu}=-\beta \kappa\opensquare T_{\mu\nu}.
\end{eqnarray}

These equations can be solved by applying the usual Green's
function method. Cancelling the surface integrals we obtain

\begin{eqnarray}
\varphi^{(1)}_{\mu\nu}=-\frac{\beta \kappa}{4\pi r}\frac{\partial^2}{\partial
t^2}{\int d^3{\bf x}^\prime T_{\mu\nu}(t-r,{\bf x}^\prime)},
\end{eqnarray}

\noindent which can also be written in the form

\begin{eqnarray}
\varphi^{(1)}_{kl}=\beta\frac{\partial^2}{\partial t^2}\varphi^{(0)}_{kl}.
\end{eqnarray}

Since $\varphi^{(1)}_{kl}$ is proportional to $\varphi^{(0)}_{kl}$
the former is also transverse and traceless.

The approximate solution up to {1th-order} is given by

\begin{eqnarray}
\label{1solu}
\varphi_{kl}^{(1)}=\varphi^{(0)}_{kl}+\beta\frac{\partial^2}{\partial
t^2}\varphi^{(0)}_{kl}.
\end{eqnarray}

Since $\varphi^{(0)}=0$, $h_{kl}^{(1)}=\varphi_{kl}^{(1)}$. Thus
we can compute the scalar amplitude, $h_s~=~\sqrt{h_{+}^2+h_{\times}^2}$, for the first quadratic
curvature order solution via equation~(\ref{1solu}). At {\it
1th-order} in $\beta$ we have

\begin{equation}
\label{corrh} {h_s}^{(1)}={\rm
h}^{(0)}+\beta\left(\frac{{h_{xx}^{(0)}}\ddot{h}_{xx}^{(0)}
+{h_{xy}^{(0)}}\ddot{h}_{xy}^{(0)}}{h^{(0)}}\right),
\end{equation}

\noindent where

\[
{h_s}^{(0)}=\sqrt{({h^{(0)}_{+}})^2+({h^{(0)}_{\times}})^2}
\]

\noindent is the scalar amplitude given by Einstein's General
Relativity and,

\[
\ddot{f}=\frac{\partial^2f}{\partial t^2}.
\]

\noindent For monochromatic signals of angular frequency $\omega $
we get the simple result

\begin{equation}
\label{corrmc}
{h_s}^{(1)}=\left  (1+\frac{\beta\omega^2}{c^2}\right) {{h_s}^{(0)}},
\end{equation}

\noindent whenever $\mid\!\!(\beta \omega^2)/c^2\!\!\mid<1$, with
$c$ being the velocity of light. The above result shows the
amplitude of a single monochromatic gravitational wave in
quadratic gravity to the first approximation in the higher
curvature terms.

\subsection{An exact solution to linearized quadratic gravity}

By adding the contributions of higher-order perturbative terms,
the solution at the {\it n-th order} can be written as

\begin{equation}
h_{kl}^{(n)}=\left(1+\frac{\beta\omega^2}{c^2}+\frac{\beta^2\omega^4}{c^4}+\frac
{\beta^3\omega^6}{c^6}+\dots+\frac{\beta^n\omega^{2n}}{c^{2n}}\right)h_{kl}^{(0)}
.
\label{seriesum}
\end{equation}

The series in $(\beta \omega^2)/c^2$ is a geometric one. This
series can be summed to give the exact result
$(1-\frac{\beta\omega^2}{c^2})^{-1}$, whenever $\mid\!\!(\beta
\omega^2)/c^2\!\!\mid<\!1$. Therefore, an exact solution for a
monochromatic wave in quadratic gravity in the linear weak field
approximation is given by

\begin{equation}
h_{kl}=\left(1-\frac{\beta\omega^2}{c^2}\right)^{-1}\!\!\! h_{kl}^{(0)},
\end{equation}

\noindent where $h_{kl}^{(0)}$ is the Einstein's General
Relativity solution.

The scalar amplitude ${h_s}$ for this solution is related to
Einstein's gravity scalar amplitude ${\rm h^{(0)}}$ by the same
factor, i. e.,

\begin{equation}
{h_s}=\left(1-\frac{\beta\omega^2}{c^2}\right)^{-1}\!\!\!{h_s}^{(0)}.
\label{quadsamp}
\end{equation}
We observe that this result is valid only for $\mid\!\!(\beta
\omega^2)/c^2\!\!\mid<\!1$ or, for waves with
frequency $0<\omega<c/\!\!\mid\!\!\beta\!\!\mid^{1/2}$.

The result derived above shows a correction to the amplitude of linear oscillations of the space-time metric due to the presence of a Ricci-squared term in gravitational Lagrangian. This result is remarkable simple and predicts a frequency dependent shift in the wave amplitude.  We point out that, in the context of higher-order gravity, this is an interesting result since it enable us to know the effect of quadratic curvature terms in linear oscillations of space-time metric by avoiding the negative energy problems that can arise due to the effect of a massive ghost, associated to the Ricci squared term, on the linearized fourth-order field equations. Since the original fourth order field equations are reduced, by the perturbative scheme, to the linearized Einstein's gravity equations in presence of an effective source, the energy density and the energy flux of gravitational waves can be computed from the same expressions for this quantities derived in Einstein's linearized gravity.     

\section{Possible implications to astrophysics and cosmology and perspectives to detectability of the amplitude correction}
\label{sec4}

Let us now discuss how the result derived in this paper can affect the predictions for the gravitational radiation generated by astrophysical and cosmological sources and discuss a possibility to detect the predicted quadratic curvature amplitude correction.  

Firstly, we need an order of magnitude estimative for the amplitude shift. To first order in $\beta$ the shift is given by the product of Einstein's linearized wave amplitude by the factor $\beta\omega^2/c^2$. Observational constraints on $\beta$ parameter derived from the bending of light by a gravitational field in the fourth-order gravity indicates an upper bound of $|\beta|~\lesssim 10^{-4}{\rm cm^2}$~\cite{accgr}. The same bound is obtained form a translation of the results of the sub-millimeter tests of the inverse square law of gravitational attraction among massive bodies~\cite{subexp}. Due to the small value of $\beta$ upper bound, the effect of the Ricci squared correction to the linearized wave amplitude will appear only for high frequency gravitational waves. Even for inspiraling compact binaries of neutron star or stellar black hole pairs, which experiments an intense growth of the frequency when the system approach the coalescence, the frequency of the gravitational oscillations cannot exceed $10^4 {\rm Hz}$. In this case, the shift in the wave amplitude will be~$\lesssim 10^{-16}h^{(0)}$, where $h^{(0)}$ is the Einstein's linearized wave amplitude. This shift is clearly non observable. 

For the amplitude shift to be observable, taking the maximal value acceptable for $\beta$, the frequency of gravitational waves must lie in the range of $10^{10}-10^{12}\;{\rm Hz}$. Compact binaries formed by pairs of mini black holes having masses $M\lesssim 4\times 10^{-5} M_\odot$ are able to emit gravitational waves with frequencies above $10^{10}\;{\rm Hz}$. As pointed out in~\cite{sublun},
a significant fraction of the dark halo of our galaxy can consist of a large number ($\sim 10^{19}$) of sub-lunar mass primordial black holes with mass $10^{-12}M_\odot\lesssim M\lesssim 10^{-7} M_\odot$. Therefore, it is natural to expect that most of these objects form binaries which can emmit very-high-frequency gravitational radiation before coalescing. Then, we can expect a background of gravitational radiation associated with these objects in the frequency band where the quadratic curvature effects might be relevant.

On the other hand, gravitational waves with frequencies in the GHz band are predicted by several models for stochastic backgrounds of gravitational waves of cosmological origin derived from inflation, string cosmology and cosmic strings scenarios~\cite{Allen,Caldw}. Then, we can expect that the result derived in the present work can affect the predictions for cosmological backgrounds of gravitational radiation, in the high (GHz) frequency spectral band. 
An indication of the effect of quadratic correction on the spectrum of a stochastic  background, which is assumed to be isotropic, stationary and unpolarized can be easily computed. The energy density of the gravitational waves is given by 
\begin{equation}
\rho_{gw}=\frac{c^2}{32\pi G}<\dot{h}_{kl}\dot{h}^{kl}>\:,
\label{edens}
\end{equation}
where $\dot{h}_{kl}$ is the time derivative of the transverse traceless gauge filed and $<>$ means spatial average over several wavelengths. If we insert the quadratic correction to gravitational waves in the equation~(\ref{edens}) and  compute the ensemble average for a stochastic background~\cite{Allen} we get the result
\begin{equation}
\Omega_{gw}(f)\equiv\frac{1}{\rho_c}\frac{d\rho_{gw}}{d(\log{f})}=\left(1+8\pi^2\frac{\beta f^2}{c^2}\right)\Omega_{gw}^{(0)}(f),
\label{Omega}
\end{equation}
where $f$ is the wave frequency ($\omega=2\pi f$), $\Omega_{gw}^{(0)}(f)$ is the ``spectral function'' predicted by Einstein's theory and $\rho_c$ is the critical energy density required to close the universe. We note that the $\Omega_{gw}$ will differ from $\Omega_{gw}^{(0)}$ only in the ($10^{10}-10^{12})\;{\rm Hz}$ frequency range. Such very-high-frequency gravitational waves reflect the behaviour of the very early universe~\cite{Allen}. We note that the expression~(\ref{Omega}) must be taken with care since the computation of $\rho_{gw}$  may be different for a given cosmological models.
Thus, a deeper inspection of how the cosmological background is affected by the quadratic curvature corrections must be carried out. 

Once very-high-frequency ($10^{10}-10^{12}\;{\rm Hz}$) gravitational radiation could be detected, this detection could be used to set up new observational constraints on the $\beta$ parameter. The gravitational wave detectors such as LIGO, VIRGO and the mass resonant detectors are not adequate  to observe such high frequency waves. The effect of quadratic curvature gravity on the linearized fluctuations of space-time metric will only be observable when a new generation of detectors with sufficient sensitivity to measure amplitudes of gravitational waves with frequencies in the range of $(10-10^3)\;{\rm GHz}$ are available. 

A gravitational wave detector which explores the effects on electromagnetic fields of the passage of gravitational waves has been proposed by Cruise~\cite{cruise}.
The proposed device can be designed to observe gravitational radiation with frequencies in the MHz and GHz range. Although the sensitivity of the Cruise detector does not reach the required sensitivity to detect gravitational wave signals, improvements in the technology might lead to more sensitive detectors in the near future. This detectors will be able to observe the gravitational radiation generated by cosmological sources at the very early moments in the universe. Therefore, the quadratic curvature corrections to gravitational waves might be detectable by a sufficiently sensitive generation of electromagnetic detectors. These detectors might also be able to observe the background radiation produced by a halo population of binary systems composed of sub-lunar mass primordial black holes, whereas these objects may actually exists.

\ack{E. C. de Rey Neto would like to thank
the Brazilian agency FAPESP for financial support (grants
00/10374-5). O. D. Aguiar and J.C.N.
de Araujo thanks the Brazilian agency CNPq for support (grants
300619/92-8, 380503/02-2, respectively).}

\end{document}